\documentclass[journal,twocolumn]{IEEEtran}   
\usepackage{graphicx}
\usepackage{epstopdf}
\usepackage{amssymb}
\usepackage{amsfonts}
\usepackage{amsmath}
\usepackage{algorithm}
\usepackage{algorithmic}
\usepackage{subeqnarray}
\usepackage{cases}
\usepackage{bm}
\usepackage{subfigure,amsmath,amssymb,cite}
\usepackage{stfloats}

\usepackage{float}
\usepackage{cuted}

\ifCLASSINFOpdf
\else
\fi
\begin{document}

\title{ Beamforming Design for RIS-Aided AF Relay Networks }

\author{Xuehui~Wang,~Feng Shu,~Riqing Chen,~Peng Zhang,~Qi Zhang,~Guiyang Xia,~Weiping shi, and~Jiangzhou~Wang,~\emph{Fellow},~\emph{IEEE}

\thanks{This work was supported in part by the National Natural Science Foundation of China (Nos.U22A2002, 62071234 and 61972093), the Major Science and Technology plan of Hainan Province under Grant ZDKJ2021022, and the Scientific Research Fund Project of Hainan University under Grant KYQD(ZR)-21008. \emph{(Corresponding authors: Feng Shu)}.}

\thanks{Xuehui Wang, Peng Zhang and Qi Zhang are with the School of Information and Communication Engineering, Hainan University,~Haikou,~570228, China.}

\thanks{Feng Shu is with the School of Information and Communication Engineering, Hainan University, Haikou 570228, China, and also with the
School of Electronic and Optical Engineering, Nanjing University of Science and Technology, Nanjing 210094, China (e-mail: shufeng0101@163.com).}

\thanks{Riqing Chen is with the Digital Fujian Institute of Big Data for Agriculture, Fujian Agriculture and Forestry University, Fuzhou 350002, China (e-mail: riqing.chen@fafu.edu.cn).}

\thanks{Guiyang Xia is with the Institute of Intelligent Agriculture, Anhui Agricultural University, Hefei 230036, China (e-mail: xiaguiyang@ahau.edu.cn).}

\thanks{Weiping Shi is with the School of Electronic and Optical Engineering, Nanjing University of Science and Technology, Nanjing 210094, China.}

\thanks{Jiangzhou Wang is with the School of Engineering, University of Kent, Canterbury CT2 7NT, U.K. (e-mail: j.z.wang@kent.ac.uk).}

}

\maketitle

\begin{abstract}

Since reconfigurable intelligent surface (RIS) is considered to be a passive reflector for rate performance enhancement, a RIS-aided amplify-and-forward (AF) relay network is presented. By jointly optimizing the beamforming matrix at AF relay and the phase shifts matrices at RIS, two schemes are put forward to address a maximizing signal-to-noise ratio (SNR) problem.
Firstly, aiming at achieving a high rate, a high-performance alternating optimization (AO) method based on Charnes-Cooper transformation and semidefinite programming (CCT-SDP) is proposed, where the optimization problem is decomposed to three subproblems solved by CCT-SDP and rank-one solutions can be recovered by Gaussian randomization.
While the optimization variables in CCT-SDP method are matrices, which leads to extremely high complexity. In order to reduce the complexity, a low-complexity AO scheme based on Dinkelbachs transformation and successive convex approximation (DT-SCA) is put forward, where matrices variables are transformed to vector variables and three decoupled subproblems are solved by DT-SCA. Simulation results verify that compared to two benchmarks (i.e. a RIS-assisted AF relay network with random phase and a AF relay network without RIS), the proposed CCT-SDP and DT-SCA schemes can harvest better rate performance. Furthermore, it is revealed that the rate of the low-complexity DT-SCA method is close to that of CCT-SDP method.

\end{abstract}


\section{Introduction}

Developing an innovative, efficient, highly reliable and resource-saving future wireless network is our vision, while with the exponential growth of smart devices and data traffic, a large and complex communication network has been formed. Meanwhile, two major practical limitations are generated, the first one is that it needs to consume a lot of energy to communicate, and the second one is that the wireless transmission channel is uncontrollable, which causes serious signal attenuation. Due to the ability to create friendly multipaths by adjusting reflection coefficient of each passive element \cite{Sf2021_1}, reconfigurable intelligent surface (RIS) is considered to have potential to improve the communication quality \cite{Tz2022,Cz2022}. Passive RIS is low-cost and low-power-consumption, which is easy to deploy \cite{Wqq2019}. Deploying properly a RIS in the communication environment, especially when the direct channel link is blocked, the transmit signal can still be received by the desired receiver for a extended coverage.

Compared with conventional relay \cite{SF2014}, such as amplify-and-forward (AF) relay and decode-and-forward (DF) relay, passive RIS has no radio frequency (RF) chain and does not need much energy to process signals \cite{Wxh2022}, thus RIS is viewed as a green reflector. A fair performance comparison between RIS and relay had been made in \cite{Gq2021,Wmx2022}, it was verified that RIS not only achieved the same energy efficiency to half-duplex relay, but also was comparable to full-duplex relay in terms of spectral efficiency in \cite{Gq2021}. Recently, RIS has become a research hot-spot and has drawn much attention. As far as we know, RIS has been widely researched in different novel communication applications, such as secure communication \cite{Jwh2021,Hs2021,Swp2021_1}, covert communication \cite{Zxb2022}, satellite communication \cite{Zbx2022,Ljw2021}, simultaneous wireless information and power transfer (SWIPT) \cite{Swp2021_2}, vehicle networks \cite{Jw2023} and spatial modulation \cite{Sf2021_2,Sf2022}.
A RIS-aided multiple input multiple output secure network was proposed in \cite{Jwh2021}, where the transmit beamforming and RIS phase shift matrix were jointly optimized by an alternate optimization (AO) algorithm for maximum sum secrecy rate. In \cite{Zbx2022}, a RIS-assisted low-earth orbit satellite system was considered. A flexible beamforming can be achieved by controlling phase shifts of RIS in a programable way, so that the time-varying channel can be cost-effectively handled. The authors applied a RIS to a traditional vehicular network in \cite{Jw2023}, where a method of maximizing instantaneous signal-to-noise ratio (SNR) was proposed to jointly optimize the transmit beamforming and RIS reflection-coefficient matrix. By evaluating the sum spectral efficiency, it was revealed that the capacity of a moving vehicular network can be enhanced by introducing a RIS.

At present, there is some existing literature researching a hybrid network including RIS and relay have appearing, which harness the benefits of RIS and relay for rate enhancement \cite{Wxh2023,Bqy2022,Kw2022}, coverage extension \cite{Az2022}, spectral efficiency and energy efficiency improvement \cite{Lgh2022,Mo2022}, and channel estimation \cite{Szw2023} of the communication network. A RIS-aided two-way DF relay wireless network was proposed in \cite{Wxh2023}, where power allocation parameters of two users and DF relay were optimized by successive convex approximation (SCA) method and maximizing determinant method to obtain maximum rate. In the hybrid network, the position of RIS plays a key role in the rate maximization problem. The authors proposed a method of optimizing RIS deployment to enhance rate in \cite{Bqy2022}, where the closed-form expressions of optimal RIS deployment were derived. Then it was verified that when RIS was close to relay, the maximum rate could be obtained. With the aim of extended coverage, three different double-RIS networks (i.e. a double-RIS network without relay, a relay-aided double-RIS network and double relay-aided double-RIS network) were respectively proposed in \cite{Az2022}, where the AO and majorization-minimization (MM) schemes were designed to optimize the double-RIS phase. Furthermore, it was demonstrated that the double relay-aided double-RIS network can achieve the higher rete than the other two networks. In \cite{Mo2022}, a multiuser downlink network aided by relay and RIS was put forward. In order to solve the optimization problem of maximizing the energy efficiency, singular value decomposition (SVD), semidefinite programming (SDP) and function approximations were respectively applied to optimize the beamforming matrices of transmitter and relay, and RIS phase shifts. In terms of channel estimation, a pilot pattern was proposed in \cite{Szw2023} to separate direct and cascaded channels for the channel estimation of a discrete-phase-shifter RIS-assisted two-way relay network, and the performance loss of finite quantization was derived.

The existing research works on hybrid networks are related to DF relay and RIS, in this case, beamforming at DF relay and phase shifts at RIS are optimized for better performance. To our best knowledge, there is few researches on such a hybrid network consisting of AF relay and RIS, which motivats us to design a RIS-aided AF relay wireless network. Aiming at improving the rate performance, two AO methods are proposed to maximize SNR. Our contributions are summarized as follow:

\begin{enumerate}

\item

A RIS-aided AF relay network is proposed, in which an non-convex optimization problem of maximizing SNR is formulated. Since the optimization variables (i.e. the beamforming matrix at AF relay and the reflecting-coefficient matrices at RIS) are coupled, a high-performance AO method based on Charnes-Cooper transformation and semidefinite programming (CCT-SDP) is proposed. To solve such a non-convex problem, vectorization and trace function is applied to convert the problem into fractional or linear optimization problem. After dropping rank-one constraints, optimization problem can be addressed by CCT and SDP algorithm. Lastly, rank-one solutions are recovered by Gaussian randomization method. Simulation results demonstrate that the proposed CCT-SDP scheme has a significant rate improvement over a RIS-assisted AF relay network with random phase and a AF relay network without RIS.

\item

The optimization variables in the proposed CCT-SDP method are matrices, which has a extremely high computational complexity (i.e. $\mathcal{O}(M^{13}+N^{6.5})$). To reduce the extremely high complexity, a low-complexity maximizing SNR scheme based on Dinkelbachs transformation and SCA (DT-SCA) is presented. By performing DT and the first-order Taylor approximation, the non-convex optimization problem are transformed to one convex problem. Since the optimization variables are vectors, the complexity of DT-SCA method (i.e. $\mathcal{O}(M^6+N^{3.5})$) is much lower than that of CCT-SDP method. Moreover, from the simulation results, it is found that the low-complexity DT-SCA scheme outperforms the existing networks with random phase and without AF relay. Additionally, its rate is approximate to that of CCT-SDP method.

\end{enumerate}

The remainder of this paper is organized as follows. We present a RIS-aided AF relay network in Section 2. In Section 3, a high-performance CCT-SDP-based AO scheme is proposed. A a low-complexity DT-SCA-based AO scheme in Section 4 is put forward. Section 5 presents the related simulation results, and Section 6 summarizes the conclusions.

\emph{Notation}: throughout this paper, we represent scalars, vectors and matrices by using letters of lower case, bold lower case, and bold upper case. The conjugate, transpose and conjugate transpose of matrix $\textbf{A}$ are expressed as $\textbf{A}^*$, $\textbf{A}^T$ and $\textbf{A}^H$. The
sign $\mathbb{E}\{\cdot\}$, $| \cdot |$, $\|\cdot\|$ and $\angle(\cdot)$ respectively represent the expectation operation, the modulus of a scalar, 2-norm and the phase of a complex number. $\text{tr}(\textbf{A})$ and $\lambda_{max}(\textbf{A})$ respectively represent the trace and the largest eigenvalue of matrix $\textbf{A}$. The sign $\textbf{I}_{N}$ stands for the N-dimensional identity matrix. $\otimes$ denotes Kronecker product.

\section{system model}
\subsection{Signal Model}
\begin{figure}[h]
\centering
\includegraphics[width=0.470\textwidth,height=0.250\textheight]{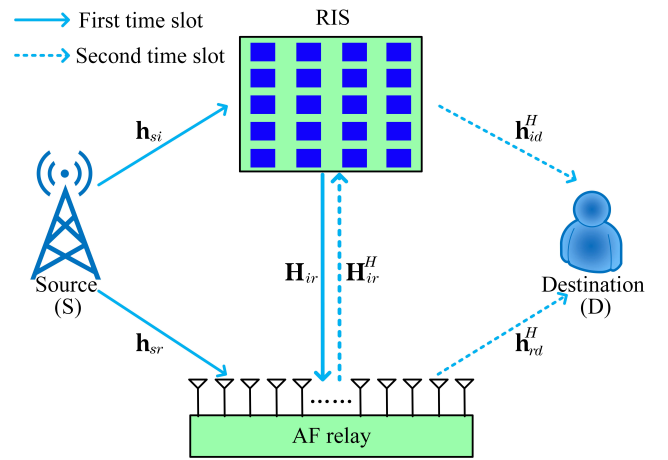}\\
\caption{A RIS-assisted AF relay wireless system.}\label{System_Model}
\end{figure}
Fig. 1 depicts a RIS-aided AF relay network, where there are a half-duplex AF relay with $M$ transmit antennas, a RIS with $N$ reflecting elements, a single-antenna source (S) and a single-antenna destination (D). Here, the network is operated in half-duplex mode. It is assumed that there is no direct path from S to D, S transmits signal towards D with the help of RIS and AF relay. The signal reflected by the RIS twice or more can be ignored because of the serious path loss, considering the RIS reflects the signal once in each time slot. Moreover, we assume all channel state informations (CSIs) are perfectly known to S, AF relay, RIS and D. In the first time slot, the received signal at AF relay is written as
\begin{equation}\label{yr_111}
\textbf{y}_r=\sqrt{P_{s}}(\textbf{h}_{sr}+\textbf{H}_{ir}\bm\Theta_1\textbf{h}_{si})x+\textbf{n}_r,
\end{equation}
where $x$ is the transmit signal from S with $\mathbb{E}[|x|^{2}]=1$, and $P_{s}$ is the total transmission power. We assume all channels follow Rayleigh
fading, $\textbf{h}_{sr}\in\mathbb{C}^{M\times 1}$, $\textbf{h}_{si}\in\mathbb{C}^{N\times 1}$ and $\textbf{H}_{ir}\in\mathbb{C}^{M\times N}$ are the channels from S to AF relay, from S to RIS and from RIS to AF relay, respectively. $\bm\Theta_1=\text{diag}(e^{j\theta_{11}},...,e^{j\theta_{1N}})$ is a diagonal reflection-coefficient matrix of RIS in the first time slot, where $\theta_{1i}\in[0,2\pi)$ is the phase shift for the $i$-th RIS reflecting element.
$\textbf{n}_{r}\in\mathbb{C}^{M\times 1}$ denotes the complex additive white Gaussian noise (AWGN) vectors at AF relay with distribution $\textbf{n}_{r}\sim\mathcal{CN}(\textbf{0}, \sigma_{r}^{2}\textbf{I}_{M})$. After receive and transmit beamforming operations are performed to AF relay, the transmit signal at AF relay can be expressed as
\begin{align}
\textbf{y}_t&=\textbf{A}\textbf{y}_{r} \nonumber\\
&=\sqrt{P_{s}}\textbf{A}(\textbf{h}_{sr}+\textbf{H}_{ir}\bm\Theta_1\textbf{h}_{si})x+\textbf{A}\textbf{n}_r,
\end{align}
where $\textbf{A}\in \mathbb{C}^{M \times M}$ represents the beamforming matrix. The transmit power at AF relay can be written as
\begin{equation}
P_r^t=P_s\|\textbf{A}(\textbf{h}_{sr} + \textbf{H}_{ir}\boldsymbol\Theta_1\textbf{h}_{si})\|^2+\|\textbf{A}\|_F^2\sigma_{r}^{2} \leq P_r,
\end{equation}
where $P_r$ is the transmit power budget at AF relay. In the second time slot, the received signal at D is given by
\begin{align}
y_d=&(\textbf{h}_{rd}^H+\textbf{h}_{id}^H\bm\Theta_2\textbf{H}_{ir}^H)\textbf{y}_t+n_d \nonumber\\
=&\sqrt{P_{s}}(\textbf{h}_{rd}^H+\textbf{h}_{id}^H\bm\Theta_2\textbf{H}_{ir}^H)\textbf{A}(\textbf{h}_{sr}+\textbf{H}_{ir}\bm\Theta_1\textbf{h}_{si})x \nonumber\\
&+(\textbf{h}_{rd}^H+\textbf{h}_{id}^H\bm\Theta_2\textbf{H}_{ir}^H)\textbf{A}\textbf{n}_r+n_d,
\end{align}
where channels $\textbf{h}_{rd}^H\in\mathbb{C}^{1\times M}$, $\textbf{h}_{id}^H\in\mathbb{C}^{1\times N}$ and $\textbf{H}_{ir}^H\in\mathbb{C}^{N\times M}$ respectively denote the links AF-D, RIS-D and AF relay-RIS. The diagonal reflection-coefficient matrix of RIS in the second time slot is expressed as $\bm\Theta_2=\text{diag}(e^{j\theta_{21}},...,e^{j\theta_{2N}})$, where $\theta_{2i}\in[0,2\pi)$ is the phase shift for the $i$-th RIS reflecting element. $n_d\in\mathbb{C}^{M\times 1}$ is the complex AWGN at D with distribution $n_d\sim\mathcal{CN}(0, \sigma_{d}^{2})$. It is assumed that $\sigma_r^2=\sigma_d^2=\sigma^2$ and $\gamma_s=\frac{P_s}{\sigma^2}$, the achievable rate can be defined as
\begin{align}
R=\frac{1}{2}\text{log}_2(1+\text{SNR}),
\end{align}
where
\begin{equation}
\text{SNR}=\frac{\gamma_s|(\textbf{h}_{rd}^H+\textbf{h}_{id}^H\bm\Theta_2\textbf{H}_{ir}^H)\textbf{A}(\textbf{h}_{sr}+\textbf{H}_{ir}\bm\Theta_1\textbf{h}_{si})|^2}
{\|(\textbf{h}_{rd}^H+\textbf{h}_{id}^H\bm\Theta_2\textbf{H}_{ir}^H)\textbf{A}\|^2+1}.
\end{equation}

\subsection{Problem Formulation}
In order to enhance the system rate performance, it is necessary to maximize SNR. The corresponding optimization problem can be formulated as
\begin{subequations}
\begin{align}
&\max \limits_{\boldsymbol\Theta_1, \boldsymbol\Theta_2, \textbf{A} } \frac{\gamma_s|(\textbf{h}_{rd}^H+\textbf{h}_{id}^H\bm\Theta_2\textbf{H}_{ir}^H)\textbf{A}(\textbf{h}_{sr}+\textbf{H}_{ir}\bm\Theta_1\textbf{h}_{si})|^2}
{\|(\textbf{h}_{rd}^H+\textbf{h}_{id}^H\bm\Theta_2\textbf{H}_{ir}^H)\textbf{A}\|^2+1} \\
&~~~\text{s.t.}~   |\boldsymbol\Theta_1(i,i)|=1,~|\boldsymbol\Theta_2(i,i)|=1, \\
&~~~~~~~~  \gamma_s\|\textbf{A}(\textbf{h}_{sr} + \textbf{H}_{ir}\boldsymbol\Theta_1\textbf{h}_{si})\|^2+\|\textbf{A}\|_F^2 \leq \gamma_r,
\end{align}
\end{subequations}
where $\gamma_r=\frac{P_r}{\sigma^2}$. By defining $\textbf{u}_1=[e^{j\theta_{11}},...,e^{j\theta_{1N}}]^T$, $\textbf{u}_2=[e^{j\theta_{21}},...,e^{j\theta_{2N}}]^H$,
$\textbf{v}_1=[\textbf{u}_1; 1]$, $\textbf{v}_2=[\textbf{u}_2; 1]$,
$\textbf{H}_{sir}=[\textbf{H}_{ir}\text{diag}(\textbf{h}_{si}),\textbf{h}_{sr}]$, and $\textbf{H}_{rid}=[\text{diag}(\textbf{h}_{id}^H)\textbf{H}_{ir}^H;\textbf{h}_{rd}^H]$,
we have
$\textbf{h}_{sr}+\textbf{H}_{ir}\bm\Theta_1\textbf{h}_{si}=\textbf{H}_{sir}\textbf{v}_1$ and
$\textbf{h}_{rd}^H+\textbf{h}_{id}^H\bm\Theta_2\textbf{H}_{ir}^H=\textbf{v}_2^H\textbf{H}_{rid}$.
Therefore, the above optimization problem can be converted to
\begin{subequations}\label{problem_1}
\begin{align}
&\max \limits_{\textbf{v}_1, \textbf{v}_2, \textbf{A} }~~ \frac{\gamma_s|\textbf{v}_2^H\textbf{H}_{rid}\textbf{A}\textbf{H}_{sir}\textbf{v}_1|^2}
{\|\textbf{v}_2^H\textbf{H}_{rid}\textbf{A}\|^2+1} \label{problem_1_1}\\
&~~\text{s.t.}~~~   |\textbf{v}_1(i)|=1,~|\textbf{v}_2(i)|=1,~\forall i=1,\cdots, N, \\
&~~~~~~~~~  \textbf{v}_1(N+1)=1,~\textbf{v}_2(N+1)=1, \\
&~~~~~~~~~  \gamma_s\|\textbf{A}\textbf{H}_{sir}\textbf{v}_1\|^2+\|\textbf{A}\|_F^2 \leq \gamma_r.\label{problem_1_2}
\end{align}
\end{subequations}
where the optimization variables $\textbf{v}_1$, $\textbf{v}_2$ and $\textbf{A}$ are coupled. To solve problem (\ref{problem_1}), two schemes: CCT-SDP-based AO method and DT-SCA-based AO method are proposed to optimize variables $\textbf{v}_1$, $\textbf{v}_2$ and $\textbf{A}$ for maximum SNR.

\section{Proposed high-performance CCT-SDP-based AO method}
Let us define $\textbf{V}_1=\textbf{v}_1\textbf{v}_1^H$ and $\textbf{V}_2=\textbf{v}_2\textbf{v}_2^H$, problem (\ref{problem_1}) can be rewritten as follows
\begin{subequations}\label{problem_2}
\begin{align}
&\max \limits_{\textbf{V}_1, \textbf{V}_2, \textbf{A} } \frac{\gamma_s\text{tr}(\textbf{V}_1\textbf{H}_{sir}^H\textbf{A}^H\textbf{H}_{rid}^H\textbf{V}_2\textbf{H}_{rid}\textbf{A}\textbf{H}_{sir})    }
{\text{tr}(\textbf{V}_2\textbf{H}_{rid}\textbf{A}\textbf{A}^H\textbf{H}_{rid}^H)+1} \label{problem_2_1}\\
&~\text{s.t.}   \textbf{V}_1(n,n)=1,\textbf{V}_2(n,n)=1, \forall n=1,\cdots, N+1, \\
&~~~~~  \gamma_s\text{tr}(\textbf{V}_1\textbf{H}_{sir}^H\textbf{A}^H\textbf{A}\textbf{H}_{sir})+\|\textbf{A}\|_F^2 \leq \gamma_r, \\
&~~~~~  \text{rank}(\textbf{V}_1)=1,~\textbf{V}_1\succeq 0, \\
&~~~~~  \text{rank}(\textbf{V}_2)=1,~\textbf{V}_2\succeq 0,
\end{align}
\end{subequations}
where the problem is non-convex with rank-one constains. By AO algorithm, the above mentioned problem (\ref{problem_2}) can be decoupled into the following three subproblems.

\subsection{Optimizing $\textbf{A}$ Given $\textbf{V}_1$ and $\textbf{V}_2$}
When $\textbf{V}_1$ and $\textbf{V}_2$ are fixed, let us define $\textbf{a}=\text{vec}(\textbf{A})\in\mathbb{C}^{M^2\times 1}$, and problem (\ref{problem_2}) can be simplified as
\begin{subequations}\label{A_0}
\begin{align}
&~\max\limits_{\textbf{a}}~~ \frac{\textbf{a}^H\textbf{B}\textbf{a}}{\textbf{a}^H\textbf{C}\textbf{a}+1} \\
&~~\text{s.t. } ~~~ \textbf{a}^H(\textbf{D}+\textbf{I}_{M^2})\textbf{a}\leq\gamma_r,
\end{align}
\end{subequations}
where matrix
$\textbf{B}=\gamma_s(\textbf{H}_{sir}^*\textbf{V}_1^*\textbf{H}_{sir}^T)\otimes(\textbf{H}_{rid}^H\textbf{V}_2\textbf{H}_{rid})$,
$\textbf{C}=\textbf{I}_M\otimes(\textbf{H}_{rid}^H\textbf{V}_2\textbf{H}_{rid})$ and
$\textbf{D}=\gamma_s(\textbf{H}_{sir}^*\textbf{V}_1^*\textbf{H}_{sir}^T)\otimes\textbf{I}_{M}$.
Upon defining
$\overline{\textbf{A}}=\textbf{a}\textbf{a}^H\in\mathbb{C}^{M^2\times M^2}$ and removing $\text{rank}(\overline{\textbf{A}})=1$ constraint, an SDR problem of problem (\ref{A_0}) is written as
\begin{subequations}\label{A_1}
\begin{align}
&~\max\limits_{\overline{\textbf{A}}}~~ \frac{\text{tr}(\textbf{B}\overline{\textbf{A}})}{\text{tr}(\textbf{C}\overline{\textbf{A}})+1}\\
&~~\text{s.t. }~~~  \text{tr}\{(\textbf{D}+\textbf{I}_{M^2})\overline{\textbf{A}}\}\leq\gamma_r, ~~\overline{\textbf{A}}\succeq 0,
\end{align}
\end{subequations}
which is a quasi-convex problem. Defining $m=(\text{tr}(\textbf{C}\overline{\textbf{A}})+1)^{-1}$, the SDR problem by using CCT operation can be transformed to the following SDP problem
\begin{subequations}\label{A_2}
\begin{align}
&~\max\limits_{\widetilde{\textbf{A}}, m}~~~ \text{tr}(\textbf{B}\tilde{\textbf{A}}) \\
&~~\text{s.t. }~~~\text{tr}(\textbf{C}\widetilde{\textbf{A}})+m=1,~~m> 0,  \\
&~~~~~~~~~~~\text{tr}\{(\textbf{D}+\textbf{I}_{M^2})\widetilde{\textbf{A}}\}\leq m\gamma_r,~~\widetilde{\textbf{A}}\succeq 0,
\end{align}
\end{subequations}
where $\widetilde{\textbf{A}}=m\overline{\textbf{A}}$. Since the objective function is concave and the constraints are convex, problem (\ref{A_2}) is convex, which can be directly solved via CVX. While constraint $\text{rank}(\overline{\textbf{A}})=1$ is ignored in the SDR problem, we apply Gaussian randomization method to achieve a solution $\overline{\textbf{A}}$ with $\text{rank}(\overline{\textbf{A}})=1$. $\textbf{a}$ can be obtained through the eigenvalue decomposition of $\overline{\textbf{A}}$, thereby, AF relay beamforming matrix $\textbf{A}$ is achieved.

\subsection{Optimizing $\textbf{V}_1$ Given $\textbf{V}_2$ and $\textbf{A}$}
Upon fixing $\textbf{V}_2$ and $\textbf{A}$, problem (\ref{problem_2}) without considering $\text{rank}(\textbf{V}_1)=1$ can be reduced to
\begin{subequations}\label{V_1}
\begin{align}
&\max \limits_{\textbf{V}_1}~~ \gamma_s\text{tr}(\textbf{V}_1\textbf{H}_{sir}^H\textbf{A}^H\textbf{H}_{rid}^H\textbf{V}_2\textbf{H}_{rid}\textbf{A}\textbf{H}_{sir}) \\
&~\text{s.t.}~~~~   \textbf{V}_1(n,n)=1, ~\forall n=1,\cdots, N+1, ~\textbf{V}_1\succeq 0,\\
&~~~~~~~~~  \gamma_s\text{tr}(\textbf{V}_1\textbf{H}_{sir}^H\textbf{A}^H\textbf{A}\textbf{H}_{sir})+\|\textbf{A}\|_F^2 \leq \gamma_r,
\end{align}
\end{subequations}
which is a standard SDP problem and efficiently solved through CVX. Similarly, rank-one solution $\textbf{V}_1$ is recovered via Gaussian randomization method.

\subsection{Optimizing $\textbf{V}_2$ Given $\textbf{V}_1$ and $\textbf{A}$}
In the case of fixing $\textbf{V}_1$, $\textbf{A}$ and taking no account of $\text{rank}(\textbf{V}_2)=1$, optimization problem (\ref{problem_2}) can be simplified as
\begin{subequations}\label{V_2}
\begin{align}
&\max \limits_{\textbf{V}_2}~~ \frac{\gamma_s\text{tr}(\textbf{V}_2\textbf{H}_{rid}\textbf{A}\textbf{H}_{sir}\textbf{V}_1\textbf{H}_{sir}^H\textbf{A}^H\textbf{H}_{rid}^H)}
{\text{tr}(\textbf{V}_2\textbf{H}_{rid}\textbf{A}\textbf{A}^H\textbf{H}_{rid}^H)+1} \\
&~\text{s.t.}~~~~   \textbf{V}_2(n,n)=1, ~\forall n=1,\cdots, N+1,~\textbf{V}_2\succeq 0.
\end{align}
\end{subequations}
Clearly, the above problem (\ref{V_2}) is similar to problem (\ref{A_1}), so that the mean of solving $\overline{\textbf{A}}$ can also be applied to seek $\textbf{V}_2$. In the same manner, the SDR problem (\ref{V_2}) can be transformed to
\begin{subequations}\label{V_2_1}
\begin{align}
&~\max\limits_{\widetilde{\textbf{V}}_2, p}~~~ \gamma_s\text{tr}(\widetilde{\textbf{V}}_2\textbf{H}_{rid}\textbf{A}\textbf{H}_{sir}\textbf{V}_1\textbf{H}_{sir}^H\textbf{A}^H\textbf{H}_{rid}^H) \\
&~~\text{s.t. }~~~  \widetilde{\textbf{V}}_2(n,n)=p, ~\forall n=1,\cdots, N+1,~\widetilde{\textbf{V}}_2\succeq 0, \\
&~~~~~~~~~~~\text{tr}(\widetilde{\textbf{V}}_2\textbf{H}_{rid}\textbf{A}\textbf{A}^H\textbf{H}_{rid}^H)+p=1,~~p> 0,
\end{align}
\end{subequations}
where $p=(\text{tr}(\textbf{V}_2\textbf{H}_{rid}\textbf{A}\textbf{A}^H\textbf{H}_{rid}^H)+1)^{-1}$. Solution $\widetilde{\textbf{V}}_2$ can be found via CVX, then Gaussian randomization method can also be applied to seek $\textbf{V}_2$ with $\text{rank}(\textbf{V}_2)=1$.

\subsection{Overall Algorithm and Complexity Analysis}
The related procedure of the proposed CCT-SDP method is summarized in Algorithm 1.
\begin{table}[h]\normalsize
\renewcommand{\arraystretch}{1}
\centering
\begin{tabular}{p{210pt}}
\hline
$\bf{Algorithm~1}$  Proposed CCT-SDP Method\\
\hline
1. Initialize $\textbf{A}^0$, $\textbf{V}_1^0$ and $\textbf{V}_2^0$, $R^0$ can be computed based on (\ref{problem_2_1}).\\
2. set $t = 0$ and convergence accuracy $\varepsilon$. \\
3.~\bf{repeat}  \\
4.   Solve problem (\ref{A_2}) to obtain $\widetilde{\textbf{A}}^{t+1}$ with given $\textbf{V}_1^t$ and $\textbf{V}_2^t$, apply Gaussian randomization to recover rank-1 solution $\overline{\textbf{A}}^{t+1}$ and obtain $\textbf{A}^{t+1}$.\\
5.   Fix $\textbf{A}^{t+1}$ and $\textbf{V}_2^t$, compute $\textbf{V}_1^{t+1}$ according to problem (\ref{V_1}), recover rank-1 solution $\textbf{V}_1^{t+1}$ via Gaussian randomization.\\
6.   Fix $\textbf{A}^{t+1}$ and $\textbf{V}_1^{t+1}$, obtain $\widetilde{\textbf{V}}_2^{t+1}$ by solving problem (\ref{V_2_1}), use Gaussian randomization to recover rank-1 solution $\textbf{V}_2^{t+1}$.\\
7.~ Compute $R^{t+1}$. \\
8.~\bf{until} \\
~~~    $\left| {R^{t+1}- R^t} \right|\le \delta$.\\
\hline
\end{tabular}
\end{table}
The corresponding total computational complexity is written as
\begin{align}\label{1_c}
&\mathcal{O}\{L_1[n_\textbf{A}\sqrt{M^2+3}(M^6+3+n_\textbf{A}(M^4+3)+n_\textbf{A}^2) \nonumber\\
& + n_{\textbf{V}_1}\sqrt{2N+3}((N+1)^3+N+2+n_{\textbf{V}_1}((N+1)^2  \nonumber\\
&+N+2) + n_{\textbf{V}_1}^2)+n_{\textbf{V}_2}\sqrt{2N+4}((N+1)^3+N+3  \nonumber\\
&+n_{\textbf{V}_2}((N+1)^2+N+3)+n_{\textbf{V}_2}^2)]\text{ln}(1/\varepsilon)\}
\end{align}
FLOPs, where $\varepsilon$ denotes the computation accuracy, $n_\textbf{A}=M^4+1$, $n_{\textbf{V}_1}=(N+1)^2$ and $n_{\textbf{V}_2}=(N+1)^2+1$ are the numbers of variables in problems (\ref{A_2}), (\ref{V_1}) and (\ref{V_2_1}), respectively. $L_1$ is the iterative number to reach the convergence condition of CCT-SDP-based AO method. Clearly, the highest order of computational complexity (\ref{1_c}) is $M^{13}$ and $N^{6.5}$ FLOPs.

\section{Proposed low-complexity DT-SCA-based AO method}

In previous section, we have proposed a high-performance CCT-SDP method to optimize the beamforming matrix $\textbf{A}$, reflecting
coefficient matrix $\textbf{V}_1$ and $\textbf{V}_2$ for a high achievable rate. While CCT-SDP method is related to the matrix optimization, which results in high complexity. Aiming at reducing the high complexity, a low-complexity DT-SCA method is proposed in this section. By optimizing one variable and fixing the other two variables, optimization problem (\ref{problem_1}) can also be decomposed to the following three subproblems.

\subsection{Optimization of $\textbf{A}$ with Fixed $\textbf{v}_1$ and $\textbf{v}_2$}
Based on Dinkelbachs transformation (DT), we reformulate the fractional optimization problem (\ref{A_0}) by introducing a slack variable $\mu$ as follow
\begin{subequations}\label{a1}
\begin{align}
&\max \limits_{ \textbf{a}, \mu} ~~~\textbf{a}^H\overline{\textbf{B}}\textbf{a}-\mu(\textbf{a}^H\overline{\textbf{C}}\textbf{a}+ 1) \label{a1_1}\\
&~\text{s.t.}~~~~~  \textbf{a}^H(\overline{\textbf{D}}+\textbf{I}_{M^2})\textbf{a}\leq\gamma_r,
\end{align}
\end{subequations}
where matrix $\overline{\textbf{B}}=\gamma_s(\textbf{H}_{sir}^*\textbf{v}_1^*(\textbf{H}_{sir}\textbf{v}_1)^T)\otimes(\textbf{H}_{rid}^H\textbf{v}_2\textbf{v}_2^H\textbf{H}_{rid})$,
$\overline{\textbf{C}}=\textbf{I}_M\otimes(\textbf{H}_{rid}^H\textbf{v}_2\textbf{v}_2^H\textbf{H}_{rid})$,
$\overline{\textbf{D}}=\gamma_s(\textbf{H}_{sir}^*\textbf{v}_1^*(\textbf{H}_{sir}\textbf{v}_1)^T)\otimes\textbf{I}_{M}$ and $\mu$ is iteratively updated in accordance with
\begin{equation}\label{mu}
\mu(t+1)=\frac{\textbf{a}^H(t)\overline{\textbf{B}}\textbf{a}(t)}{\textbf{a}^H(t)\overline{\textbf{C}}\textbf{a}(t)+1},
\end{equation}
where $t$ is the iteration index. It is worthy noting that problem (\ref{a1}) is non-convex because of the non-concave objective function (\ref{a1_1}). Here, we apply SCA method to approximate $\textbf{a}^H\overline{\textbf{B}}\textbf{a}$ to a linear function given by
\begin{equation}
\textbf{a}^H\overline{\textbf{B}}\textbf{a} \geq 2\Re\{\textbf{a}^H\overline{\textbf{B}}\widetilde{\textbf{a}}\}-\widetilde{\textbf{a}}^H\overline{\textbf{B}}\widetilde{\textbf{a}}.
\end{equation}
The above inequation is the corresponding first-order Taylor expansion of $\textbf{a}^H\overline{\textbf{B}}\textbf{a}$ at feasible point $\widetilde{\textbf{a}}$. Therefore, problem (\ref{a1}) can be transformed to
\begin{subequations}\label{a2}
\begin{align}
&\max \limits_{ \textbf{a}} ~~~2\Re\{\textbf{a}^H\overline{\textbf{B}}\widetilde{\textbf{a}}\}-\widetilde{\textbf{a}}^H\overline{\textbf{B}}\widetilde{\textbf{a}}-
\mu(\textbf{a}^H\overline{\textbf{C}}\textbf{a}+ 1) \label{a2_1}\\
&~\text{s.t.}~~~~~  \textbf{a}^H(\overline{\textbf{D}}+\textbf{I}_{M^2})\textbf{a}\leq\gamma_r.
\end{align}
\end{subequations}
Due to the fact that $\mu$ is nondecreasing, the convergence of the objective function (\ref{a2_1}) can be guaranteed. Obviously, problem (\ref{a2}), consisting of concave objective function and convex constraint, is convex, and its solution $\textbf{a}$ can be computed via CVX. Consequently, $\textbf{A}$ can be obtained.

\subsection{Optimization of $\textbf{v}_1$ with Fixed $\textbf{v}_2$ and $\textbf{A}$}
Given $\textbf{v}_2$ and $\textbf{A}$, problem (\ref{problem_1}) can be reduced to
\begin{subequations}\label{v1}
\begin{align}
&\max \limits_{\textbf{v}_1 }~~~~ \textbf{v}_1^H\textbf{E}\textbf{v}_1 \label{v1_1}\\
&~\text{s.t.}~~~~~~  |\textbf{v}_1(i)|=1,~\forall i=1,\cdots, N, \label{v1_2}\\
&~~~~~~~~~~~  \textbf{v}_1(N+1)=1, \label{v1_3}\\
&~~~~~~~~~~~  \textbf{v}_1^H\textbf{F}\textbf{v}_1+\|\textbf{A}\|_F^2 \leq \gamma_r, \label{v1_4}
\end{align}
\end{subequations}
where matrix
$\textbf{E}=\gamma_s\textbf{H}_{sir}^H\textbf{A}^H\textbf{H}_{rid}^H\textbf{v}_2\textbf{v}_2^H\textbf{H}_{rid}\textbf{A}\textbf{H}_{sir}$ and
$\textbf{F}=\gamma_s\textbf{H}_{sir}^H\textbf{A}^H\textbf{A}\textbf{H}_{sir}$.
The above problem is non-convex because of a non-concave objective function (\ref{v1_1}) and a unit-modulus constraint (\ref{v1_2}). (\ref{v1_1})
can be approximated as a linear function through its first-order Taylor expansion at feasible vector $\widetilde{\textbf{v}}_1$, which is
\begin{equation}
\textbf{v}_1^H\textbf{E}\textbf{v}_1 \geq 2\Re\{\textbf{v}_1^H\textbf{E}\widetilde{\textbf{v}}_1\}-\widetilde{\textbf{v}}_1^H\textbf{E}\widetilde{\textbf{v}}_1.
\end{equation}
The unit-modulus constraint (\ref{v1_2}) can be relaxed to
\begin{equation}\label{v1_5}
|\textbf{v}_1(i)|^2\leq1,~\forall i=1,\cdots, N.
\end{equation}
After the above transformation, we have the following optimization problem
\begin{subequations}\label{v1_6}
\begin{align}
&\max \limits_{\textbf{v}_1 }~~~~ 2\Re\{\textbf{v}_1^H\textbf{E}\widetilde{\textbf{v}}_1\}-\widetilde{\textbf{v}}_1^H\textbf{E}\widetilde{\textbf{v}}_1 \\
&~\text{s.t.}~~~~~~  \text{(\ref{v1_3})},~\text{(\ref{v1_4})},~\text{(\ref{v1_5})}.
\end{align}
\end{subequations}
The above problem is a convex optimization problem, and its solution is denoted as $\widehat{\textbf{v}}_1$ achieved via CVX. Thereby, the solution to problem (\ref{v1}) is given by
\begin{equation}
\textbf{v}_1=e^{j\angle{\frac{\widehat{\textbf{v}}_1}{\widehat{\textbf{v}}_1(N+1)}}}.
\end{equation}

\subsection{Optimization of $\textbf{v}_2$ with Fixed $\textbf{v}_1$ and $\textbf{A}$}
When $\textbf{v}_1$ and $\textbf{A}$ are given, problem (\ref{problem_1}) can be simplified as
\begin{subequations}\label{v2}
\begin{align}
&~\max\limits_{\textbf{v}_2}~~~~   \frac{\textbf{v}_2^H\textbf{G}\textbf{v}_2}{\textbf{v}_2^H\textbf{J}\textbf{v}_2}\\
&~~\text{s.t.}~~~~~~  |\textbf{v}_2(i)|=1,~\forall i=1,\cdots, N, \\
&~~~~~~~~~~~~  \textbf{v}_2(N+1)=1,\label{v2_1}
\end{align}
\end{subequations}
where matrix
$\textbf{G}=\gamma_s\textbf{H}_{rid}\textbf{A}\textbf{H}_{sir}\textbf{v}_1\textbf{v}_1^H\textbf{H}_{sir}^H\textbf{A}^H\textbf{H}_{rid}^H$ and
$\textbf{J}=\textbf{H}_{rid}\textbf{A}\textbf{A}^H\textbf{H}_{rid}^H+\frac{\textbf{I}_{N+1}}{N+1}$.
Similarly, the SCA method is also applied to achieve the low bound of (\ref{v2_1}). Its first-order Taylor expansion at feasible vector $\widetilde{\textbf{v}}_2$ can be expressed as \cite{Gxr2020}
\begin{equation}
\frac{\textbf{v}_2^H\textbf{G}\textbf{v}_2}{\textbf{v}_2^H\textbf{J}\textbf{v}_2} \geq 2\Re\{\textbf{g}^H\textbf{v}_2\}+d,
\end{equation}
where
\begin{subequations}
\begin{align}
\textbf{g}^H=&\frac{\widetilde{\textbf{v}}_2^H\textbf{G}}{\widetilde{\textbf{v}}_2^H\textbf{J}\widetilde{\textbf{v}}_2}
-\widetilde{\textbf{v}}_2^H( \textbf{J}-\lambda_{max}(\textbf{J})\textbf{I}_{N+1} )
\frac{\widetilde{\textbf{v}}_2^H\textbf{G}\widetilde{\textbf{v}}_2}{(\widetilde{\textbf{v}}_2^H\textbf{J}\widetilde{\textbf{v}}_2)^2}, \\ d=&-[2\lambda_{max}(\textbf{J})(N+1)-\widetilde{\textbf{v}}_2^H\textbf{J}\widetilde{\textbf{v}}_2]
\frac{\widetilde{\textbf{v}}_2^H\textbf{G}\widetilde{\textbf{v}}_2}{(\widetilde{\textbf{v}}_2^H\textbf{J}\widetilde{\textbf{v}}_2)^2}.
\end{align}
\end{subequations}
In the same manner, problem (\ref{v2}) can be transformed to
\begin{subequations}\label{v2_2}
\begin{align}
&~\max\limits_{\textbf{v}_2}~~~~   2\Re\{\textbf{g}^H\textbf{v}_2\}+d \\
&~~\text{s.t.}~~~~~~  |\textbf{v}_2(i)|^2 \leq 1,~\forall i=1,\cdots, N,~\text{(\ref{v2_1})},
\end{align}
\end{subequations}
which is convex. Its optimal solution $\widehat{\textbf{v}}_2$ can be efficiently found by CVX. Therefore, the solution to problem (\ref{v2}) can be denoted as
\begin{equation}
\textbf{v}_2=e^{j\angle{\frac{\widehat{\textbf{v}}_2}{\widehat{\textbf{v}}_2(N+1)}}}.
\end{equation}


\subsection{Overall Algorithm and Complexity Analysis}
The related procedure of the proposed DT-SCA method is summarized in Algorithm 2.
\begin{table}[h]\normalsize
\renewcommand{\arraystretch}{1}
\centering
\begin{tabular}{p{210pt}}
\hline
$\bf{Algorithm~2}$  Proposed DT-SCA Method\\
\hline
1. Initialize $\textbf{A}^0$, $\textbf{v}_1^0$ and $\textbf{v}_2^0$. Achieve $R^0$ in line with (\ref{problem_1}). Set $t = 0$ and convergence accuracy $\varepsilon$. \\
2.~\bf{repeat}  \\
3.~~   Given $\textbf{v}_1^t$ and $\textbf{v}_2^t$, find $\textbf{a}^{t+1}$ by solving problem (\ref{a2}), and obtain $\textbf{A}^{t+1}$.\\
4.~   Given $\textbf{A}_1^{t+1}$ and $\textbf{v}_2^t$, find $\textbf{v}_1^{t+1}$ by solving problem (\ref{v1_6}).\\
5.~~   Given $\textbf{A}^{t+1}$ and $\textbf{v}_1^{t+1}$, find $\textbf{v}_2^{t+1}$ by solving problem (\ref{v2_2}).\\
6.~~   Achieve $R^{t+1}$. \\
7.~\bf{until} \\
~~~    $\left| {R^{t+1}- R^t} \right|\le \delta$.\\
\hline
\end{tabular}
\end{table}
The corresponding total complexity of Algorithm 2 is
\begin{align}\label{2_c}
&\mathcal{O}\{L_2[n_\textbf{a}\sqrt{2}(M^4+n_\textbf{a}^2) \nonumber\\
& + n_{\textbf{v}_1}\sqrt{2N+3}(n_{\textbf{v}_1}+N+1+(N+2)^2  \nonumber\\
& + n_{\textbf{v}_1}^2)+n_{\textbf{v}_2}\sqrt{2N+1}(n_{\textbf{v}_2}+N+1+n_{\textbf{v}_2}^2)]\text{ln}(1/\varepsilon)\}
\end{align}
FLOPs. The number of variables $n_\textbf{a}$, $n_{\textbf{v}_1}$ and $n_{\textbf{v}_2}$ are respectively equal to $M^2$, $N+1$ and $N+1$ in problems (\ref{a2}), (\ref{v1_6}) and (\ref{v2_2}). $L_2$ is the iterative number satisfied the convergence condition of DT-SCA method. The highest order of computational complexity (\ref{2_c}) is $M^6$ and $N^{3.5}$ FLOPs, which is much lower than that of CCT-SDP method.

\section{Simulation and Numerical Results}
Numerical simulations are applied to verify the rate performance of the proposed two schemes in this section. For convenience, the positions of S, D, RIS and AF relay are respectively set as (0, 0, 0), (0, 100m, 0), ($-$10m, 50m, 20m) and (10m, 50m, 10m) in three-dimensional (3D) space. According to $PL(d)=PL_0-10{\alpha}\text{log}_{10}(\frac{d}{d_0})$, the path loss between transmitter and receiver can be computed. $\alpha$ represents the path loss exponent, $d$ is the distance, and $PL_0= -$30dB denotes the reference path loss at $d_0=1$m. Here, the path loss exponents of channel link S-RIS, S-AF relay, RIS-AF relay, RIS-D and AF relay-D are respectively set as 2.0, 3.5, 2.0, 2.0 and 3.5. Additionally, let $P_r=$30dBm, $\sigma^2=-$90dBm.

\begin{figure}[h]
\centering
\includegraphics[width=0.400\textwidth,height=0.240\textheight]{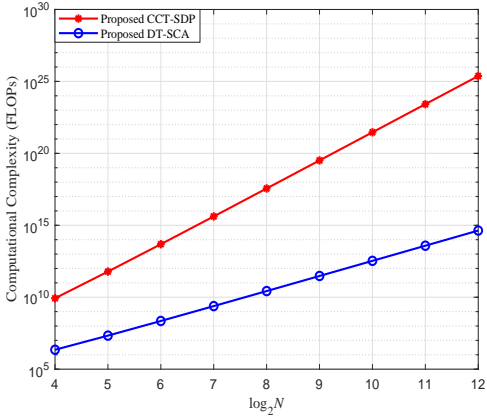}\\
\caption{  Computational complexity versus $N$ with ($M, L_1, L_2$) $=$ (2, 10, 10).  }\label{figure2_Complexity}
\end{figure}
Fig. 2 shows the computational complexity of the proposed two methods versus $N$ with ($M, L_1, L_2$) $=$ (2, 10, 10). It verifies that as N increases, the computational complexity of the proposed CCT-SDP method and DT-SCA method increase. Meanwhile, it is obvious that the complexity of CCT-SDP method is far higher than that of DT-SCA method, which is consistent with our analysis of the complexity in previous sections.

\begin{figure}[h]
\centering
\includegraphics[width=0.400\textwidth,height=0.240\textheight]{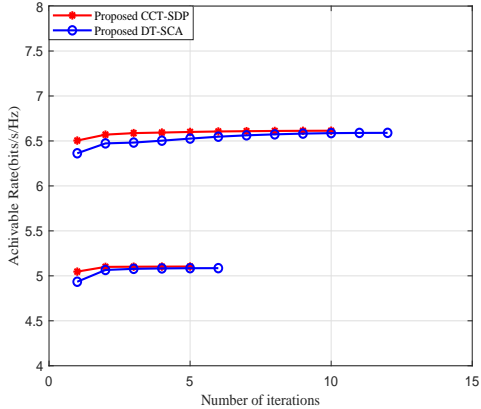}\\
\caption{   Convergence of proposed methods with ($M, N, P_r$) $=$ (2, 256, 30dBm).  }\label{figure3_Iteration}
\end{figure}
Fig. 3 verifies the convergence of the proposed two algorithms with $P_s =$ 15dBm and 25dBm. When $P_s =$ 15dBm, it needs about four iterations to be convergent for the proposed CCT-SDP and DT-SCA methods. In the case of $P_s =$ 25dBm, the proposed two methods can come up to maximum rate within eight iterations. It is revealed that the proposed CCT-SDP and DT-SCA methods are convergent and feasible.

\begin{figure}[h]
\centering
\includegraphics[width=0.400\textwidth,height=0.240\textheight]{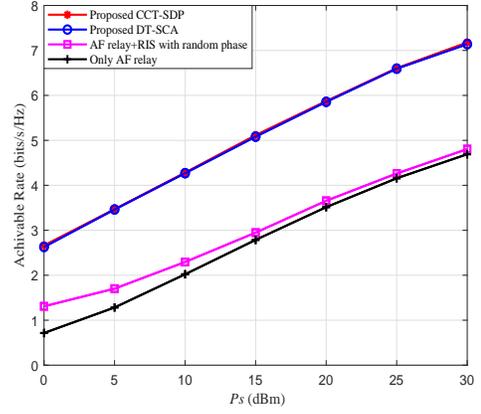}\\
\caption{  Achievable rate versus $P_s$ with ($M, N, P_r$) $=$ (2, 256, 30dBm).  }\label{figure4_Rate_Vs_Ps}
\end{figure}
Fig. 4 depicts the achievable rate versus $P_s$ with (M, N, $P_r$)= (2, 256, 30dBm). Obviously, the rate performance of the proposed CCT-SDP and DT-SCA methods increase as the transmit power $P_s$ at S increase. Since multipaths have been created with the aid of RIS, and beamforming matrix at AF relay and phase shifts at RIS have been optimized, the proposed two schemes can perform much better than a RIS-assisted AF relay network with random phase and a AF relay network without RIS. Moreover, the rate of DT-SCA method is slightly lower than that of CCT-SDP scheme in the whole $P_s$ region.

\begin{figure}[h]
\centering
\includegraphics[width=0.400\textwidth,height=0.240\textheight]{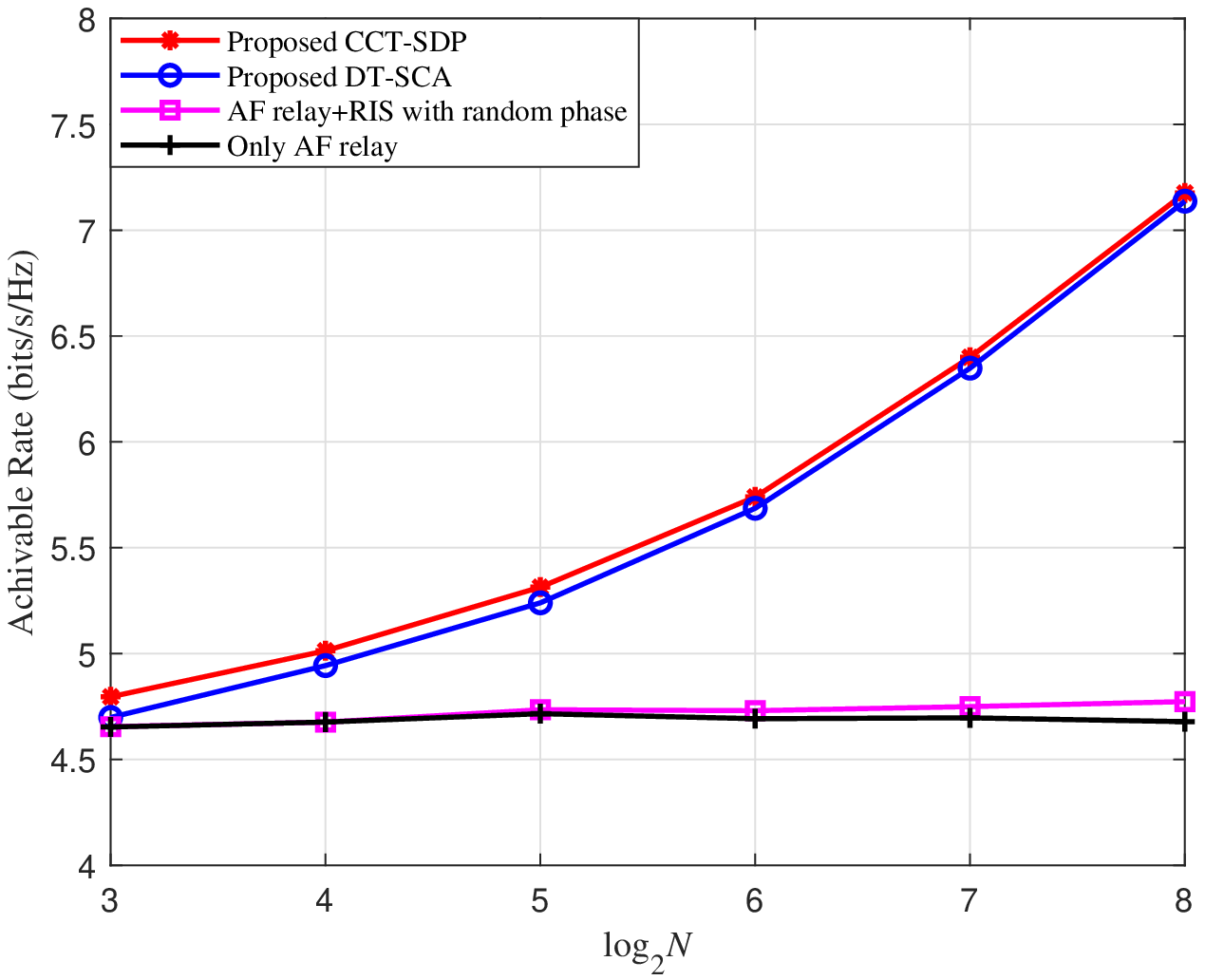}\\
\caption{  Achievable rate versus $N$ with ($M, P_s, P_r$) $=$ (2, 30dBm, 30dBm).  }\label{figure5_Rate_Vs_N}
\end{figure}
Fig. 5 shows the achievable rate versus the number $N$ of RIS elements with ($M, P_s, P_r$) $=$ (2, 30dBm, 30dBm). For small-scale RIS, the gap between the proposed two methods (i.e. CCT-SDP and DT-SCA) and two benchmark schemes  (i.e. a RIS-assisted AF relay network with random phase and a AF relay network without RIS) is small. As the number $N$ of RIS elements increases, the rates of CCT-SDP and DT-SCA methods increase, which widen the gap gradually. For medium-scale and large-scale RIS, the gap between CCT-SDP and DT-SCA method is growing smaller. When $N=$ 256, the rate of DT-SCA method is approximate to that of CCT-SDP method. Meanwhile, compared to the two benchmark schemes, a 50\% rate gain can be obtained by the proposed two schemes.

\section{Conclusion}

In this paper, a RIS-aided AF relay wireless network was investigated. There were two AO methods based on the criterion of Max SNR proposed, one is CCT-SDP and the other is DT-SCA. Here, the beamforming matrix at AF relay and phase shifts matrices at RIS were jointly optimized for rate performance enhancement. From simulation, it was verified that the proposed CCT-SDP and DT-SCA schemes are convergent, which can obtain an apparent rate improvement compared to a RIS-assisted AF relay network with random phase and a AF relay network without RIS. Besides that, it was proved that the rate performance achieved by the low-complexity DT-SCA method is slightly lower than that of the high-performance CCT-SDP method.

\ifCLASSINFOpdf
\else
\fi

\bibliographystyle{IEEEtran}
\bibliography{reference}

\begin{thebibliography}{10}
\providecommand{\url}[1]{#1}
\csname url@samestyle\endcsname
\providecommand{\newblock}{\relax}
\providecommand{\bibinfo}[2]{#2}
\providecommand{\BIBentrySTDinterwordspacing}{\spaceskip=0pt\relax}
\providecommand{\BIBentryALTinterwordstretchfactor}{4}
\providecommand{\BIBentryALTinterwordspacing}{\spaceskip=\fontdimen2\font plus
\BIBentryALTinterwordstretchfactor\fontdimen3\font minus
  \fontdimen4\font\relax}
\providecommand{\BIBforeignlanguage}[2]{{%
\expandafter\ifx\csname l@#1\endcsname\relax
\typeout{** WARNING: IEEEtran.bst: No hyphenation pattern has been}%
\typeout{** loaded for the language `#1'. Using the pattern for}%
\typeout{** the default language instead.}%
\else
\language=\csname l@#1\endcsname
\fi
#2}}
\providecommand{\BIBdecl}{\relax}
\BIBdecl

\bibitem{Sf2021_1}
F.~Shu, Y.~Teng, J.~Li, M.~Huang, W.~Shi, J.~Li, Y.~Wu, and J.~Wang, ``Enhanced
  secrecy rate maximization for directional modulation networks via {IRS},''
  \emph{IEEE Trans. Commun.}, vol.~69, no.~12, pp. 8388--8401, Dec. 2021.

\bibitem{Tz2022}
Z.~Tian, Z.~Chen, M.~Wang, Y.~Jia, L.~Dai, and S.~Jin, ``Reconfigurable
  intelligent surface empowered optimization for spectrum sharing: Scenarios
  and methods,'' \emph{IEEE Trans. Veh. Technol.}, vol.~17, no.~2, pp. 74--82,
  June 2022.

\bibitem{Cz2022}
Z.~Chen, J.~Tang, X.~Y. Zhang, D.~K.~C. So, S.~Jin, and K.-K. Wong, ``Hybrid
  evolutionary-based sparse channel estimation for irs-assisted mmwave mimo
  systems,'' \emph{IEEE Trans. Wireless Commun.}, vol.~21, no.~3, pp.
  1586--1601, Mar. 2022.

\bibitem{Wqq2019}
Q.~Wu and R.~Zhang, ``Intelligent reflecting surface enhanced wireless network
  via joint active and passive beamforming,'' \emph{IEEE Trans. Wireless
  Commun.}, vol.~18, no.~11, pp. 5394--5409, Nov. 2019.

\bibitem{SF2014}
F.~Shu, Y.~Lu, Y.~Chen, X.~You, J.~Wang, M.~Wang, W.~Sheng, and Q.~Chen,
  ``High-sum-rate beamformers for multi-pair two-way relay networks with
  amplify-and-forward relaying strategy,'' \emph{Sci. China Inf. Sci.},
  vol.~57, pp. 1--11, Feb. 2014.

\bibitem{Wxh2022}
X.~Wang, F.~Shu, W.~Shi, X.~Liang, R.~Dong, J.~Li, and J.~Wang, ``Beamforming
  design for {IRS}-aided decode-and-forward relay wireless network,''
  \emph{IEEE Trans. Green Commun. Netw.}, vol.~6, no.~1, pp. 198--207, Mar.
  2022.

\bibitem{Gq2021}
Q.~Gu, D.~Wu, X.~Su, J.~Jin, Y.~Yuan, and J.~Wang, ``Performance comparisons
  between reconfigurable intelligent surface and full/half-duplex relays,'' in
  \emph{IEEE 94th Veh. Technol. Conf. (VTC2021-Fall)}, Dec. 2021, pp. 01--06.

\bibitem{Wmx2022}
M.~Wang, W.~Duan, G.~Zhang, M.~Wen, J.~Choi, and P.-H. Ho, ``On the achievable
  capacity of cooperative {NOMA} networks: {RIS} or relay?'' \emph{IEEE
  Wireless Commun. Lett.}, vol.~11, no.~8, pp. 1624--1628, Aug. 2022.

\bibitem{Jwh2021}
W.~Jiang, B.~Chen, J.~Zhao, Z.~Xiong, and Z.~Ding, ``Joint active and passive
  beamforming design for the {IRS}-assisted {MIMOME-OFDM} secure
  communications,'' \emph{IEEE Trans. Veh. Technol.}, vol.~70, no.~10, pp.
  10\,369--10\,381, Oct. 2021.

\bibitem{Hs2021}
S.~Hong, C.~Pan, H.~Ren, K.~Wang, K.~K. Chai, and A.~Nallanathan, ``Robust
  transmission design for intelligent reflecting surface-aided secure
  communication systems with imperfect cascaded {CSI},'' \emph{IEEE Trans.
  Wireless Commun.}, vol.~20, no.~4, pp. 2487--2501, Apr. 2021.

\bibitem{Swp2021_1}
W.~Shi, J.~Li, G.~Xia, Y.~Wang, X.~Zhou, Y.~Zhang, and F.~Shu, ``Secure
  multigroup multicast communication systems via intelligent reflecting
  surface,'' \emph{China Commun.}, vol.~18, no.~3, pp. 39--51, Mar. 2021.

\bibitem{Zxb2022}
X.~Zhou, S.~Yan, Q.~Wu, F.~Shu, and D.~W.~K. Ng, ``Intelligent reflecting
  surface ({IRS})-aided covert wireless communications with delay constraint,''
  \emph{IEEE Trans. Wireless Commun.}, vol.~21, no.~1, pp. 532--547, Jan. 2022.

\bibitem{Zbx2022}
B.~Zheng, S.~Lin, and R.~Zhang, ``Intelligent reflecting surface-aided {LEO}
  satellite communication: Cooperative passive beamforming and distributed
  channel estimation,'' \emph{IEEE J. Sel. Areas Commun.}, vol.~40, no.~10, pp.
  3057--3070, Oct. 2022.

\bibitem{Ljw2021}
J.~Lee, W.~Shin, and J.~Lee, ``Performance analysis of {IRS}-assisted {LEO}
  satellite communication systems,'' in \emph{Int. Conf. Inf. and Commun.
  Technol. Conv. (ICTC)}, Dec. 2021, pp. 323--325.

\bibitem{Swp2021_2}
W.~Shi, X.~Zhou, L.~Jia, Y.~Wu, F.~Shu, and J.~Wang, ``Enhanced secure wireless
  information and power transfer via intelligent reflecting surface,''
  \emph{IEEE Commun. Lett.}, vol.~25, no.~4, pp. 1084--1088, Apr. 2021.

\bibitem{Jw2023}
W.~Jiang and H.~D. Schotten, ``Intelligent reflecting vehicle surface: A novel
  {IRS} paradigm for moving vehicular networks,'' in \emph{IEEE Mil. Commun.
  Conf. (MILCOM)}, Jan. 2023, pp. 793--798.

\bibitem{Sf2021_2}
F.~Shu, X.~Jiang, X.~Liu, L.~Xu, G.~Xia, and J.~Wang, ``Precoding and transmit
  antenna subarray selection for secure hybrid spatial modulation,'' \emph{IEEE
  Trans. Wireless Commun.}, vol.~20, no.~3, pp. 1903--1917, Mar. 2021.

\bibitem{Sf2022}
F.~Shu, L.~Yang, X.~Jiang, W.~Cai, W.~Shi, M.~Huang, J.~Wang, and X.~You,
  ``Beamforming and transmit power design for intelligent reconfigurable
  surface-aided secure spatial modulation,'' \emph{IEEE J. Sel. Topics Signal
  Process.}, vol.~16, no.~5, pp. 933--949, Aug. 2022.

\bibitem{Wxh2023}
X.~Wang, P.~Zhang, F.~Shu, W.~Shi, and J.~Wang, ``Power allocation for
  irs-aided two-way decode-and-forward relay wireless network,'' \emph{IEEE
  Trans. Veh. Technol.}, vol.~72, no.~1, pp. 1337--1342, Jan. 2023.

\bibitem{Bqy2022}
Q.~Bie, Y.~Liu, Y.~Wang, X.~Zhao, and X.~Y. Zhang, ``Deployment optimization of
  reconfigurable intelligent surface for relay systems,'' \emph{IEEE Trans.
  Green Commun. Netw.}, vol.~6, no.~1, pp. 221--233, Mar. 2022.

\bibitem{Kw2022}
W.~Khalid, M.~Shahjalal, and H.~Yu, ``Outage performance analysis of hybrid
  relay-reconfigurable intelligent surface networks,'' in \emph{27th Asia Pac.
  Conf. on Commun. (APCC)}, Nov. 2022, pp. 253--254.

\bibitem{Az2022}
Z.~Abdullah, S.~Kisseleff, K.~Ntontin, W.~A. Martins, S.~Chatzinotas, and
  B.~Ottersten, ``Double-{RIS} communication with {DF} relaying for coverage
  extension: Is one relay enough?'' in \emph{IEEE Int. Conf. Commun. (ICC)},
  Aug. 2022, pp. 2639--2644.

\bibitem{Lgh2022}
G.-H. Li, D.-W. Yue, S.-N. Jin, and Q.~Hu, ``Hybrid double-{RIS} and {DF}-relay
  for outdoor-to-indoor communication,'' \emph{IEEE Access}, vol.~10, pp.
  126\,651--126\,663, Dec. 2022.

\bibitem{Mo2022}
M.~Obeed and A.~Chaaban, ``Joint beamforming design for multiuser {MISO}
  downlink aided by a reconfigurable intelligent surface and a relay,''
  \emph{IEEE Trans. Wireless Commun.}, vol.~21, no.~10, pp. 8216--8229, Oct.
  2022.

\bibitem{Szw2023}
Z.~Sun, X.~Wang, S.~Feng, X.~Guan, F.~Shu, and J.~Wang, ``Pilot optimization
  and channel estimation for two-way relaying network aided by {IRS} with
  finite discrete phase shifters,'' \emph{IEEE Trans. Veh. Technol.}, pp. 1--6,
  2023.

\bibitem{Gxr2020}
X.~Guan, Q.~Wu, and R.~Zhang, ``Joint power control and passive beamforming in
  {IRS}-assisted spectrum sharing,'' \emph{IEEE Commun. Lett.}, vol.~24, no.~7,
  pp. 1553--1557, July 2020.

\end{thebibliography}

\end{document}